\documentclass[12pt]{article}
\usepackage{graphicx,fancyhdr}

\begin{document}
%\begin{abstract}
%\end{abstract}
%\tableofcontents
\def\K{\mathord{\cal K}}
\def\la{\langle}
\def\ra{\rangle}
\def\ltsim{\mathop{\,<\kern-1.05em\lower1.ex\hbox{$\sim$}\,}}
\def\gtsim{\mathop{\,>\kern-1.05em\lower1.ex\hbox{$\sim$}\,}}

\begin{center}

%Draft: 4 (29.10.2002)

\subsection*{ Many-body Green's
function theory for thin ferromagnetic anisotropic Heisenberg films:
treatment of the exchange anisotropy}
\vspace{2cm}

P. Fr\"obrich$^+$, and P.J. Kuntz
\vspace{1cm}

Hahn-Meitner-Institut Berlin, Glienicker Stra{\ss}e 100, D-14109 Berlin,
Germany,\\
$^+$also: Institut f\"ur Theoretische Physik, Freie Universit\"at Berlin\\
Arnimallee 14, D-14195 Berlin, Germany\\
\end{center}
\vspace{3cm}

{\bf Abstract.}
The many-body Green's function theory developed in our previous work for
treating the reorientation of the magnetization of thin ferromagnetic films
is extended to include the exchange anisotropy. This leads to additional
momentum dependencies which require some non-trivial changes in the formalism.
The theory
is developed for arbitrary spin values S and for multilayers.
The effects of the exchange anisotropy and the single-ion
anisotropy, which was treated in our earlier work, on the magnetic
properties of thin ferromagnetic films are compared.
\vspace{2cm}

{\bf PACS.} 75.10Jm Quantized spin models - 75.30Ds Spin waves - 75.70Ak
Magnetic properties of monolayers and thin films

\newpage
\subsection*{1. Introduction}
There is increasing activity in experimental and theoretical investigations of
thin magnetic films and multilayers. Of particular interest is the
reorientation of the magnetization as function of temperature and film
thickness. The simplest approach for treating thin ferromagnetic films is the
application of mean field theory (MFT) to a Heisenberg model, by either
diagonalizing the corresponding single-particle Hamiltonian \cite{Mos94}
or applying thermodynamic perturbation theory \cite{Jen96}. This approximation
completely neglects collective excitations (magnons = spin waves).

In order to take the influence of these collective excitations into account, we
have turned to many-body Green's function theory (GFT), which allows reliable
calculations over the entire range of temperature of interest. In reference
\cite{EFJK99} we treated a spin $S=1/2$ Heisenberg monolayer in a magnetic
field
and found, by comparing with Quantum Monte Carlo (QMC) calculations, that a
Tyablikov (RPA) \cite{Tya59} decoupling is a very good approximation.
Therefore,
we did not try to go beyond RPA in the subsequent paper \cite{FJK00}, in
which
we treated the field-induced magnetic reorientation of a ferromagnetic $S=1$
monolayer, whereby a second-order single-ion anisotropy was also included.
Whereas the exchange
interaction terms are decoupled by RPA, this is not allowed for the terms
coming from the single-ion anisotropy because this leads to unphysical results.
Instead, we applied a decoupling procedure  proposed by
Anderson and Callen \cite{AC64}, which, however, is a good approximation only
for
small anisotropies. This was shown to be the case in reference \cite{FKS02},
where
we were able to treat the single-ion anisotropy exactly (for any strength)
by introducing higher-order Green's functions and subsequently taking advantage
of relations between products of spin operators, which leads to an automatic
closure of the hierarchy of the equations of motion for the Green's functions
with respect to the anisotropy terms; the terms from the exchange interaction
are still decoupled  by a generalized RPA scheme. In reference \cite{HFKTJ02}
we have
investigated the quality of this approach by comparing with QMC calculations.
In reference \cite{FKS02} we treated the spin $S=1$ case only; the
formal generalization to spins $S>1$ is possible, but its numerical realisation
is quite cumbersome. This is not the case when remaining at the level of the
lowest-order Green's functions and applying the Anderson-Callen decoupling.
In this case,
not only is the treatment of spins $S>1$ feasible, but also multilayers can
be
described, as was done in reference \cite{FJKE00}. To make the
treatment of
multilayers practicable, we had to apply a new method which not only uses the
eigenvalues but also the eigenvectors of the non-symmetric matrix which governs
the equations of motion for the Green's functions. We mention a few related
papers in which Green's function theory is also applied to  spin
reorientation problems. In reference \cite{Eric91} Green's
functions are applied to the reorientation problem after a
Holstein-Primakoff mapping to bosons, which is only a valid description at low
temperatures. In reference \cite{Guo00} the reorientation is
obtained by the competition of an approximately decoupled
single-ion anisotropy and a schematic shape anisotropy. In \cite{HLT99} the
interplay of the dipole coupling with an easy-plane single-ion anisotropy is
discussed, and in \cite{DGS02} the in-plane dipole coupling anisotropy of a
square ferromagnetic Heisenberg monolayer is considered.

In our previous work mentioned above, we treated an isotropic Heisenberg
exchange interaction plus a single-ion anisotropy, the magnetic dipole
coupling, and an external magnetic field. In the present paper, we include the
exchange anisotropy in the formalism and discuss its effect on the magnetic
properties of thin ferromagnetic films, in particular in comparison to the
influence of the single-ion anisotropy. As will be shown, the treatment of the
exchange anisotropy requires some non-trivial changes in the formalism, which
are necessary due to additional momentum dependencies which are absent in the
treatment of the single-ion anisotropies.

The paper is organized as follows.
In Section 2 we establish the Green's function formalism.
For pedagogical reasons,
a large part of the formalism is demonstrated for the monolayer case and
subsequently generalised to the multilayer case, which can easily be done. In
Section 3, we discuss the field-induced reorientation of the magnetization when
using the exchange anisotropy for determining the orientation at temperature
$T=0$. We fix the strength of the exchange anisotropy in such a way that it
produces the same Curie temperature for a monolayer as the application of the
single-ion anisotropy strength used in our previous work \cite{FJK00,FJKE00}.
This enables
a comparison between the roles of the single-ion and exchange anisotropies on
the magnetic properties of thin ferromagnetic films. Section 4 contains a
summary of the results.

\subsection*{2. The Green's function formalism}

We formulate the theory in such a way that the results of our previous work
\cite{FJK00,FJKE00} are obtained as limiting cases.

We consider a spin Hamiltonian consisting of an isotropic Heisenberg exchange
interaction with strength $J_{kl}$ between nearest neighbour lattice sites,
an exchange
anisotropy with strength $D_{kl}$, a second-order single-ion lattice
anisotropy with strength $K_{2,k}$, a
magnetic dipole coupling  with strength $g_{kl}$, and an external magnetic
field ${\bf B}=(B^x,B^y,B^z)$:
\begin{eqnarray}
{\cal
H}=&-&\frac{1}{2}\sum_{<kl>}J_{kl}(S_k^-S_l^++S_k^zS_l^z)
-\frac{1}{2}\sum_{<kl>}D_{kl}S_k^zS_l^z-\sum_kK_{2,k}(S_k^z)^2\nonumber\\
&-&\sum_k\Big(\frac{1}{2}B^-S_k^++\frac{1}{2}B^+S_k^-+B^zS_k^z\Big)\nonumber\\
&+&\frac{1}{2}\sum_{kl}\frac{g_{kl}}{r_{kl}^5}\Big(r_{kl}^2(S_k^-S_l^++S_k^zS_l
^ z )-3({\bf S}_k{\bf r}_{kl})({\bf S}_l{\bf r}_{kl})\Big) .
\label{5.1}
\end{eqnarray}
Here the notation $S_k^{\pm}=S_k^x\pm iS_k^y$ and $B^{\pm}=B^x\pm iB^y$ is
introduced, where $k$ and $l$ are lattice site indices and $<kl>$ indicates
summation over nearest neighbours only. The only difference from
reference \cite{FJKE00} is the additional exchange-anisotropy term.

In order to treat the reorientation problem for general spin $S$, we need the
following Green's functions
\begin{equation}
G_{ij,\eta}^{\alpha,mn}(\omega)=\la\la
S_i^\alpha;(S_j^z)^m(S_j^-)^n\ra\ra_{\omega,\eta}\ ,
\label{5.2}
\end{equation}
where $\alpha=(+,-,z)$ takes care of all directions in space, $\eta=\pm 1$
refers to the anticommutator or commutator Green's functions, respectively, and
$n\geq 1, m\geq 0$ are positive integers, necessary for dealing with higher
spin values $S$.

The exact equations of motion are
\begin{equation}
\omega G_{ij,\eta}^{\alpha,mn}(\omega)=A_{ij,\eta}^{\alpha,mn}+\la\la
[S_i^\alpha,{\cal H}]_-;(S_j^z)^m(S_j^-)^n\ra\ra_{\omega,\eta}
\label{5.3}
\end{equation}
with the inhomogeneities
\begin{equation}
A_{ij,\eta}^{\alpha,mn}=\la[S_i^\alpha,(S_j^z)^m(S_j^-)^n]_{\eta}\ra,
\label{5.4}
\end{equation}
where $\la ...\ra=Tr(...e^{-\beta{\cal H}})$. The equations, leaving out for the moment the terms due to
the dipole coupling, are given explicitly by
\begin{eqnarray}
\omega G_{ij,\eta}^{\pm,mn}&=&A_{ij,\eta}^{\pm,mn}\nonumber\\
& &\mp\sum_{k}J_{ik}\Big(\la\la S_i^zS_k^\pm;(S_j^z)^m(S_j^-)^n\ra\ra
-\la\la S_k^zS_i^\pm;(S_j^z)^m(S_j^-)^n\ra\ra\Big)\nonumber\\
& &\pm\sum_k D_{ik}\la\la S_k^zS_i^\pm;(S_j^z)^m(S_j^-)^n\ra\ra\nonumber\\
& &\pm K_{2,i}\la\la(S_i^\pm
S_i^z+S_i^zS_i^\pm);(S_j^z)^m(S_j^-)^n\ra\ra\nonumber\\
& &\mp B^\pm G_{ij,\eta}^{z,mn}\pm B^zG_{ij,\eta}^{\pm,mn}\nonumber\\
\omega G_{ij,\eta}^{z,mn}&=&A_{ij(\eta)}^{z,mn}\nonumber\\
& &+\frac{1}{2}\sum_kJ_{ik}\la\la(S_i^-S_k^+-S_k^-S_i^+);
(S_j^z)^m(S_j^-)^n\ra\ra\nonumber\\
& &-\frac{1}{2}B^- G_{ij,\eta}^{+,mn}+\frac{1}{2}B^+G_{ij,\eta}^{-,mn}.
\label{5.5}
\end{eqnarray}

After solving these equations the components of the magnetization can be
determined from the Green's functions via the spectral theorem.
A solution is possible by establishing a closed system of equations by
decoupling the higher-order Green's functions on the right hand sides.
Contrary to reference \cite{FKS02}, where we proceed to higher-order Green's
functions, we stay here at the level of the lowest-order equations.
For the exchange-interaction and exchange-anisotropy terms, we use a
generalized Tyablikov- (or RPA-) decoupling
\begin{equation}
\la\la S_i^\alpha S_k^\beta;(S_j^z)^m(S_j^-)^n\ra\ra_\eta \simeq\la
S_i^\alpha\ra
G_{kj,\eta}^{\beta,mn}+\la S_k^\beta\ra G_{ij,\eta}^{\alpha,mn} .
\label{5.6}
\end{equation}
The terms from the single-ion anisotropy have to be decoupled differently,
because an RPA decoupling leads to unphysical results; e.g. for spin $S=1/2$,
the terms due to the single-ion anisotropy do not vanish in RPA, as they should
do, because in this case $\sum_i K_{2,i}\la (S_i^z)^2\ra$ is a constant and
should not influence the equations of motion.
In the appendix of ref. \cite{FJK00} we investigated different decoupling
schemes proposed in the literature, e.g. those of Lines \cite{Lin67} or that of
Anderson and Callen \cite{AC64}, which should be reasonable for single-ion
anisotropies small compared to the exchange interaction. We found the
Anderson-Callen decoupling to be
most adequate. It consists in implementing the suggestion of Callen
\cite{Cal63} to improve the RPA by treating the diagonal terms arising from
the single-ion anisotropy as well. This leads to
\begin{eqnarray}
& &\la\la(S_i^\pm S_i^z+S_i^zS_i^\pm);(S_j^z)^m(S_j^-)^n\ra\ra_\eta \nonumber\\
& &\simeq 2\la S_i^z\ra\Big(1-\frac{1}{2S^2}[S(S+1)-\la
S_i^zS_i^z\ra]\Big)G_{ij,\eta}^{\pm,mn}.
\label{5.7}
\end{eqnarray}
This term vanishes for $S=1/2$ as it should.

After a Fourier transform to momentum space, one obtains, for a
ferromagnetic film with $N$ layers,
$3N$ equations of motion for a $3N$-dimensional Green's function vector ${\bf
G}^{mn}$:
\begin{equation}
(\omega{\bf 1}-{\bf \Gamma}){\bf G}^{mn}={\bf A}^{mn},
\label{5.8}
\end{equation}
where ${\bf 1}$ is the
$3N\times 3N$ unit matrix. The Green's function vectors and inhomogeneity
vectors each
consist of $N$  three-dimensional subvectors which are characterized by the
layer indices $i$ and $j$

\begin{equation}
{\bf G}_{ij}^{mn}({\bf{k},\omega})\  =
\left( \begin{array}{c}
G_{ij}^{+,mn}({\bf{k}},\omega) \\ G_{ij}^{-,mn}({\bf{k}},\omega)  \\
G_{ij}^{z,mn}({\bf{k}},\omega)
\end{array} \right), \hspace{0.5cm}
{\bf A}_{ij}^{mn} {=}
 \left( \begin{array}{c} A_{ij}^{+,mn} \\ A_{ij}^{-,mn} \\
A_{ij}^{z,mn} \end{array} \right) \;.
\label{5.9} \end{equation}

The equations of motion are then expressed in terms of these layer vectors, and
$3\times 3 $ submatrices ${\bf \Gamma}_{ij}$ of the $3N\times 3N$
matrix ${\bf\Gamma}$
\begin{equation}
\left[ \omega {\bf 1}-\left( \begin{array}{cccc}
{\bf\Gamma}_{11} & {\bf\Gamma}_{12} & \ldots & {\bf\Gamma}_{1N} \\
{\bf\Gamma}_{21} & {\bf\Gamma}_{22} & \ldots & {\bf\Gamma}_{2N} \\
\ldots & \ldots & \ldots & \ldots \\
{\bf\Gamma}_{N1} & {\bf\Gamma}_{N2} & \ldots & {\bf\Gamma}_{NN}
\end{array}\right)\right]\left[ \begin{array}{c}
{\bf G}_{1j} \\ {\bf G}_{2j} \\ \ldots \\ {\bf G}_{Nj} \end{array}
\right]=\left[ \begin{array}{c}
{\bf A}_{1j}\delta_{1j} \\ {\bf A}_{2j}\delta_{2j} \\ \ldots \\
{\bf A}_{Nj}\delta_{Nj} \end{array}
\right] \;, \hspace{0.5cm} j=1,...,N\;.
\label{5.10}
\end{equation}
After applying the decoupling procedures  (\ref{5.6}) and (\ref{5.7}),
the  ${\bf \Gamma}$ matrix reduces to a band matrix with zeros in the
${\bf \Gamma}_{ij}$ sub-matrices, when $j>i+1$ and $j<i-1$.
The  diagonal sub-matrices ${\bf \Gamma}_{ii}$ are of size $3\times 3$
and have the form
\begin{equation}
 {\bf \Gamma}_{ii}= \left( \begin{array}
{@{\hspace*{3mm}}c@{\hspace*{5mm}}c@{\hspace*{5mm}}c@{\hspace*{3mm}}}
\;\;\;H^z_i & 0 & -H^+_i \\ 0 & -H^z_i & \;\;\;H^-_i \\
-\frac{1}{2}\tilde{H}^-_i & \;\frac{1}{2}\tilde{H}^+_i & 0
\end{array} \right)
\ . \label{5.11}
\end{equation}
where
\begin{eqnarray}
H^z_i&=&Z_i+\la S_i^z\ra J_{ii}(q-\gamma_{\bf k})\ ,
\nonumber\\
Z_i&=&B^z_i
+D_{ii}q\la S^z_i\ra
+(J_{i,i+1}+D_{i,i+1})\la S_{i+1}^{z}\ra+(J_{i,i-1}+D_{i,i-1})\la
S_{i-1}^{z}\ra\nonumber\\
& &+K_{2,i}2\la S_i^z\ra
\Big(1-\frac{1}{2S^2}[S(S+1)-\la S_i^zS_i^z\ra]\Big)\ ,
\nonumber \\
\tilde{H}^\pm_i&=&B^\pm_i+\la S_i^\pm\ra J_{ii}(q-\gamma_{\bf
k})
+J_{i,i+1}\la S_{i+1}^{\pm}\ra+J_{i,i-1}\la
S_{i-1}^{\pm}\ra \ ,
\nonumber\\
H^\pm_i&=&\tilde{H}^\pm_i-\la S_i^\pm\ra D_{ii}\gamma_{\bf k} \ .
\label{5.12}
\end{eqnarray}
For a square lattice and a lattice constant taken to be unity,  $\gamma_{\bf
k}=2(\cos k_x+\cos k_y)$, and $q=4$ is the
number of intra-layer nearest neighbours. Except for the exchange-anisotropy
terms, $D_{ij}$, these equations are the same as in reference \cite{FJKE00}.
Putting
all $D_{ij}=0$, one has $\tilde{H}_i^\pm=H_i^\pm$. Note that owing to
the momentum dependence in $H_i^\pm$ coming from the exchange anisotropy,
 $\tilde{H}_i^\pm\neq H_i^\pm$, which
forbids a naive extension of the formalism of reference \cite{FJKE00}, as
discussed below.

Approximating the dipole coupling by mean field theory (MFT), which is
a good
approximation when the dipole coupling strength is small as compared to the
exchange interaction strength ( as proved in appendix A of \cite{FJKE00}), one
finds that the effects of the dipole coupling can be included as an effective
field: %
\begin{eqnarray}
B_i^\pm&=&B^\pm +\sum_{j=1}^N g_{ij}\la S_j^\pm\ra T^{|i-j|},\nonumber\\
B_i^z&=&B^z-2\sum_{j=1}^N g_{ij}\la S_j^z\ra T^{|i-j|},
\label{5.13}
\end{eqnarray}
where the lattice sums for a two-dimensional square lattice are given by
($n=|n-j|$)
\begin{equation}
T^n=\sum_{lm}\frac{l^2-n^2}{(l^2+m^2+n^2)^{5/2}}.
\label{5.14}
\end{equation}
The indices $lm$ run over all sites of the $j^{th}$ layer excluding terms with
$l^2+m^2+n^2=0$.

One observes that the dipole coupling in MFT leads to a renormalization of the
external field: there is an enhancement of the transverse fields, and a
reduction of the field perpendicular to the film.

The $3\times 3$
off-diagonal sub-matrices ${\bf \Gamma}_{ij}$ for $j= i\pm 1$ are of the
form %
\begin{equation}
 {\bf \Gamma}_{ij} = \left( \begin{array}
{@{\hspace*{3mm}}c@{\hspace*{5mm}}c@{\hspace*{5mm}}c@{\hspace*{3mm}}}
-J_{ij}\la S_i^z\ra & 0 & \;\;\;(J_{ij}+D_{ij})\la S_i^+\ra \\
0 & \;\;J_{ij}\la S_i^z\ra & -(J_{ij}+D_{ij)}\la S_i^-\ra \\
\frac{1}{2}J_{ij}\la S_i^-\ra &
-\frac{1}{2}J_{ij}\la S_i^+\ra & 0 \end{array} \right) \;.
\label{5.15}
\end{equation}

When treating the monolayer, one can use the spectral theorem for calculating
the components of the magnetization. This was done in reference
\cite{FJK00} for the case of spin $S=1$ and the single-ion anisotropy by using
the commutator Green's functions.
In order to obtain sufficient equations it was necessary, to add equations
coming
from the condition that the commutator Green's functions have to be regular at
$\omega=0$, which we call the regularity conditions.

For the multilayer problem, however, a naive application of the spectral
theorem turned
out to be forbiddingly difficult. Instead we invented a method, which we call
the eigenvector method, that uses the eigenvectors as well as  the
eigenvalues of the
${\Gamma}$-matrix governing the equations of motion. This opened up
a practicable way to treat multilayers \cite{FJKE00}.

If instead, anticommutator Green's functions are used, it is not
necessary to introduce the regularity conditions, which nevertheless are valid.
We demonstrate this explicitly for the monolayer.
The use of the anticommutator Green's functions
also suggests a way for finding the procedure which can deal
with the additional ${\bf k}$-dependencies coming from the exchange anisotropy.

For simplicity, we consider the reorientation in the $x-z$-plane, i.e.
we use as external field ${\bf
B}=(B^x,0,B^z)$. The equations of motion for the monolayer in this case are
\begin{equation}
 \left( \begin{array}
{@{\hspace*{3mm}}c@{\hspace*{5mm}}c@{\hspace*{5mm}}c@{\hspace*{3mm}}}
\;\;\;\omega-H^z   &0 & H^x \\ 0 &\omega +H^z & \;\;\;-H^x \\
+\frac{1}{2}\tilde{H}^x & \;-\frac{1}{2}\tilde{H}^x & \omega \end{array}
 \right)
\left( \begin{array}{l}
G^{+,mn}_{{\bf k},\eta}\\
G^{-,mn}_{{\bf k},\eta}\\
G^{z,mn}_{{\bf k},\eta}
\end{array}\right)
=
\left( \begin{array}{l}
A^{+,mn}_{{\bf k},\eta}\\
A^{-,mn}_{{\bf k},\eta}\\
A^{z,mn}_{{\bf k},\eta}
\end{array}\right).
\label{5.16}
\end{equation}
This system of equations has three eigenvalues
\begin{equation}
\omega_1=0;\ \ \ \omega_{2,3}=\pm\epsilon_{\bf
k}=\pm\sqrt{H^zH^z+\tilde{H}^xH^x} .
\label{5.17}
\end{equation}
and the equations are solved by
\begin{equation}
G^{\alpha,mn}_{{\bf k},\eta}=\frac{\Delta^{\alpha,mn}_\eta}{\Delta},
\label{5.18}
\end{equation}
where $\Delta^{\alpha,mn}_\eta$ is the determinant of the matrix in
equation (\ref{5.16}) where column
$\alpha $ is replaced by the inhomogeneity vector, and
$\Delta=\omega(\omega-\epsilon_{\bf k})(\omega+\epsilon_{\bf k})$.

Now the spectral theorem \cite{GHE01} is applied in momentum
space
\begin{equation}
C_{\bf k}^{\alpha,mn}=\la (S^z)^m(S^-)^nS^\alpha\ra_{\bf k}=
\lim_{\delta\rightarrow 0}\frac{i}{2\pi}\int_{-\infty}^\infty
\frac{d\omega}{e^{\beta\omega}+1}\Big(G_{{\bf
k},\eta=+1}^{\alpha,mn}(\omega+i\delta)-G_{{\bf
k},\eta=+1}^{\alpha,mn}(\omega+i\delta)\Big) .
\label{5.19}
\end{equation}
Using the relation between anticommutator and commutator
\begin{equation}
A_{{\bf k},\eta=+1}^{\alpha,mn}=A_{\eta=-1}^{\alpha,mn}+2C_{\bf k}^{\alpha,mn},
\label{5.20}
\end{equation}
where it is important that the commutator inhomogeneities
$A_{\eta=-1}^{\alpha,mn}$ do not depend on the momentum ${\bf k}$,
one obtains the following set of equations
\begin{eqnarray}
H^zC_{\bf k}^{+,mn}-H^xC_{\bf
k}^{z,mn}=A_{\eta=-1}^{+,mn}\Big(\frac{1}{2}\epsilon_{\bf
k}\coth(\frac{\beta\epsilon_{\bf
k}}{2})-\frac{1}{2}H^z\Big)+\frac{1}{2}H^xA_{\eta=-1}^{z,mn},
\label{5.21}\\
-H^zC_{\bf k}^{-,mn}+H^xC_{\bf
k}^{z,mn}=A_{\eta=-1}^{-,mn}(\frac{1}{2}\epsilon_{\bf
k}\coth(\frac{\beta\epsilon_{\bf k
}}{2})+\frac{1}{2}H^z)-\frac{1}{2}H^xA_{\eta=-1 }^{z,mn},
\label{5.22}\\
\tilde{H}^xC_{\bf k}^{+,mn}-\tilde{H}^xC_{\bf
k}^{-,mn}=\frac{1}{2}\tilde{H}^x(A_{\eta=-1}^{-,mn}
-A_{\eta=-1}^{+,mn})-
\epsilon_{\bf
k}\coth(\frac{\beta\epsilon_{\bf k}}{2})A_{\eta=-1}^{z,mn}.
\label{5.23}
\end{eqnarray}
Because the observable correlations are in real space,
we have to perform a corresponding Fourier transformation
$\la (S^z)^m_i(S^-)^n_iS^\alpha_i\ra =C_i^{\alpha,mn}=\frac{1}{N}\sum_{\bf
k}C_{\bf k}^{\alpha,mn}$.

Fourier transform of equation (\ref{5.22}) yields
\begin{equation}
-C_i^{-,mn}+\frac{1}{N}\sum_{\bf k}\frac{H^x}{H^z}C_{\bf k}^{z,mn}=
\frac{1}{2}A_{\eta=-1}^{-,mn}+
\frac{1}{2}A_{\eta=-1}^{-,mn}\frac{1}{N}\sum_{\bf k}\frac{\epsilon_{\bf
k}}{H^z}\coth(\frac{\beta\epsilon_{\bf k}}{2})
-\frac{1}{2}A_{\eta=-1}^{z,mn}\frac{1}{N}\sum_{\bf k}\frac{H^x}{H^z}.
\label{5.24}
\end{equation}
Putting this into the Fourier transform of equation (\ref{5.21}),
one can eliminate
the term $\frac{1}{N}\sum_{\bf k}\frac{H^x}{H^z}C_{\bf k}^{z,mn}$.
This will turn out to be important in the later discussion of the
eigenvector method where the formalism has to be modified because
one cannot take the ${\bf k}$-dependent terms outside the sum
(integral).
One obtains
\begin{equation}
C_i^{+,mn}-C_i^{-,mn}-\frac{1}{2}(A_{\eta=-1}^{-,mn}-A_{\eta=-1}^{+,mn})=
\frac{1}{2}(A_{\eta=-1}^{-,mn}+A_{\eta=-1}^{+,mn})
\frac{1}{N}\sum_{\bf k}\frac{\epsilon_{\bf
k}}{H^z}\coth(\frac{\beta\epsilon_{\bf k}}{2}) .
\label{5.25}
\end{equation}
The Fourier transform of equation (\ref{5.23}) can be done directly and  gives
\begin{equation}
C_i^{+,mn}-C_i^{-,mn}-\frac{1}{2}(A_{\eta=-1}^{-,mn}-A_{\eta=-1}^{+,mn})
=-A_{\eta=-1}^{z,mn}
\frac{1}{N}\sum_{\bf k}\frac{\epsilon_{\bf
k}}{\tilde{H}^x}\coth(\frac{\beta\epsilon_{\bf k}}{2}) .
\label{5.26}
\end{equation}
Equations (\ref{5.25}) and (\ref{5.26}) are sufficient to determine the
observables.

To elucidate these equations  we derive the explicit expressions for
spins $S=1/2$ and $S=1$.

For $S=1/2$ we need $m=0$ and $n=1$. This gives two equations of motion which
determine $\la S^x\ra$ and $\la S^z\ra$.

From equation (\ref{5.25}) one finds with $\la S_i^-S_i^+\ra=1/2-\la S_i^z\ra$
and $\la S_i^-S_i^-\ra=0$ and $A_{\eta=-1}^{-,01}=0$ and
$A_{\eta=-1}^{+,10}=2\la S_i^z\ra$
\begin{equation}
\frac{1}{2}=\la S^z\ra \frac{1}{N}\sum_{\bf k}\frac{\epsilon_{\bf
k}}{H^z}\coth(\frac{\beta\epsilon_{\bf k}}{2}),
\label{5.27}
\end{equation}
and from equation (\ref{5.26}) with $A_{\eta=-1}^{z,10}=-\la S_i^-\ra=-\la
S_i^x\ra$
\begin{equation}
\frac{1}{2}=\la S^x\ra \frac{1}{N}\sum_{\bf k}\frac{\epsilon_{\bf
k}}{\tilde{H}^x}\coth(\frac{\beta\epsilon_{\bf k}}{2}).
\label{5.28}
\end{equation}
This are two equations which determine the two unknowns $\la S^x\ra$ and
$\la S^z\ra$. No regularity conditions are necessary.

For $S=1$ one needs equations (\ref{5.25}) and (\ref{5.26}) not only for
$(n=1, m=0)$ but also for $(n=1, m=1), (n=2, m=0), (n=3, m=0)$.
This yields 8 equations for the eight unknowns
$\la S^-\ra, \la S^z\ra, \la S^-S^-\ra, \la S^zS^-\ra, \la S^zS^z\ra, $
$\la S^zS^zS^-\ra, \la S^zS^zS^-\ra, \la S^-S^-S^zS^z\ra$.

The left sides of equations (\ref{5.25}) and (\ref{5.26}) are the same.
Therefore we write them pairwise.
For $(n=1,m=0)$ we have
\begin{equation}
2-\la S^zS^z\ra-\la S^-S^-\ra=\left\{
\begin{array}{l}
\la S^z\ra\frac{1}{N}\sum_{\bf k}\frac{\epsilon_{\bf
k}}{H^z}\coth(\frac{\beta\epsilon_{\bf k}}{2})\\
\\
\la S^-\ra\frac{1}{N}\sum_{\bf k}\frac{\epsilon_{\bf
k}}{\tilde{H}^x}\coth(\frac{\beta\epsilon_{\bf k}}{2}).
 \end{array}\right.
\label{5.29}
\end{equation}
For $(n=1, m=1)$ and $\la S^zS^zS^z\ra=\la S^z\ra$, valid for $S=1$, we have
\begin{eqnarray}
& &\frac{1}{2}(\la S^z+\la S^zS^z\ra -\la S^-S^-\ra-2\la S^zS^-S^-\ra
-2)=\nonumber\\
& &\left\{
\begin{array}{l}\frac{1}{2}\Big(\la S^-S^-\ra+3\la S^zS^z\ra-\la S^z\ra-2\Big)
\frac{1}{N}\sum_{\bf k}\frac{\epsilon_{\bf
k}}{H^z}\coth(\frac{\beta\epsilon_{\bf k}}{2})\\
\\
\la S^zS^-\ra
\frac{1}{N}\sum_{\bf k}\frac{\epsilon_{\bf
k}}{\tilde{H}^x}\coth(\frac{\beta\epsilon_{\bf k}}{2}).
 \end{array}\right.
\label{5.30}
\end{eqnarray}

For $(n=2, m=0)$ and $\la S^-S^-S^-\ra=0$ we find
\begin{eqnarray}
& &3\la S^-\ra-\la S^-S^z\ra-\la S^-S^zS^z\ra +2\la S^zS^-\ra
=\nonumber\\
& &\left\{
\begin{array}{l}(2\la S^zS^-\ra+\la S^-\ra)
\frac{1}{N}\sum_{\bf k}\frac{\epsilon_{\bf
k}}{H^z}\coth(\frac{\beta\epsilon_{\bf k}}{2})\\
\\
2\la S^-S^-\ra
\frac{1}{N}\sum_{\bf k}\frac{\epsilon_{\bf
k}}{\tilde{H}^x}\coth(\frac{\beta\epsilon_{\bf k}}{2}).
 \end{array}\right.
\label{5.31}
\end{eqnarray}
For $(n=3, m=0)$ we have with $\la S^-S^-S^-S^-\ra=0$ and $\la S^-S^-S^-\ra=0$
\begin{eqnarray}
& &2\la S^-S^-\ra-\la S^-S^-S^z\ra-\la S^-S^-S^zS^z\ra
=\nonumber\\
& &\left\{
\begin{array}{l}(3\la S^zS^-S^-\ra+3\la S^-S^-\ra)
\frac{1}{N}\sum_{\bf k}\frac{\epsilon_{\bf
k}}{H^z}\coth(\frac{\beta\epsilon_{\bf k}}{2})\\
\\
0.
 \end{array}\right.
\label{5.32}
\end{eqnarray}
From equations (\ref{5.32}) we find
$\la S^-S^-S^zS^z\ra=2\la S^-S^-\ra-\la S^-S^-S^z\ra$ and
$\la S^zS^-S^-\ra=-\la S^-S^-\ra$.
The remaining correlations are determined by the previous six equations.

Instead of using these 8 equations one can also apply the regularity
conditions, as we did in reference \cite{FJK00} when working with the
commutator Green's functions. Then one can express all
correlations in terms of the correlations
$\la S^z\ra $ and $\la S^zS^z\ra$ in the case of spin $S=1$, and in terms of
the moments $\la (S^z)^n\ra$ with (n=1,..., 2S) for arbitrary spin S, and one
need only solve two equations in the $S=1$ case or $2S$ equations for
arbitrary spin values $S$.

The regularity conditions are obtained from the fact that the commutator
Green's function has to be regular at the origin
\begin{equation}
\lim_{\omega\rightarrow 0} \omega G_{{\bf k},\eta=-1}^{\alpha,mn}=0,
\label{5.33}
\end{equation}
which leads to the relations
\begin{equation}
\tilde{H}^xA_{\eta=-1}^{+,mn}+\tilde{H}^xA_{\eta=-1}^{-,mn}+2H^zA_{\eta=-1}^{z,
m n } =0.
\label{5.34}
\end{equation}
Note that these relations are also
obtained by equating equations (\ref{5.25}) and (\ref{5.26}).

For $m=0, n=1$
we obtain the first regularity condition
\begin{equation}
\la S^x\ra=\frac{\tilde{H}^x}{H^z}\la S^z\ra;
\label{5.35}
\end{equation}
i.e. the knowledge of $\la S^z\ra$ determines $\la S^x\ra$.

From the definitions (\ref{5.12}) one sees that the prefactor is momentum
independent and the relation generalized to the multilayer can be written as
\begin{equation}
\la S^x_i\ra=\frac{\tilde{H}^x_i}{H^z_i}\la S^z_i\ra=\frac{B^x_i+J_{i,i+1}\la
S_{i+1}^x\ra +J_{i,i-1}\la S_{i-1}^x \ra}{Z_i}\la S^z_i\ra.
\end{equation}

In the case of spin $S=1$, equation (\ref{5.25}) with $(n=1, m=0)$ and $(n=1,
m=1)$ together with the
regularity conditions  determines all desired correlations for spin $S=1$.
We demonstrate this by deriving the equations for $\la S^z\ra$ and
$\la S^zS^z\ra$  already derived with the commutator Green's functions in
reference \cite{FJK00}.

Equation (\ref{5.25}) gives for $(n=1, m=0)$ with $\la S^-S^+\ra=2-\la
S^z\ra-\la S^zS^z\ra$
\begin{equation}
2-\la S^zS^z\ra-\la S^-S^-\ra=\la S^z\ra\frac{1}{N}\sum_{{\bf
k}}\sqrt{1+\frac{\tilde{H}^xH^x}{H^zH^z}}\coth (\frac{\beta\epsilon_{\bf
k}}{2}).
\label{5.36}
\end{equation}
From the regularity conditions (\ref{5.34}) one finds for S=1
\begin{equation}
\la
S^-S^-\ra=\frac{(\frac{\tilde{H}^x}{H^z})^2}{2-(\frac{\tilde{H}^x}{H^z})^2}
(3\la S^zS^z\ra-2) .
\label{5.37}
\end{equation}
Putting this into equation (\ref{5.36}) gives the first of the desired
equations
\begin{equation}
4-2\la S^zS^z\ra\Big(1+(\frac{\tilde{H}^x}{H^z})^2\Big)- \la
S^z\ra\Big(2-(\frac{\tilde{H}^x}{H^z})^2\Big) \frac{1}{N}\sum_{{\bf
k}}\sqrt{1+\frac{\tilde{H}^xH^x}{H^zH^z}}\coth (\frac{\beta\epsilon_{\bf
k}}{2})=0.
\label{5.38}
\end{equation}
The second  equation is obtained from equation (\ref{5.25}) with
$(n=1, m=1)$, and
the regularity conditions which relates $\la S^zS^-S^-\ra$ to $\la S^z\ra$
and $\la S^zS^z\ra$. This leads to
\begin{equation}
\la S^z\ra \Big(2-(\frac{\tilde{H}^x}{H^z})^2\Big)-2\Big(3\la S^zS^z\ra-2\Big)
\frac{1}{N}\sum_{{\bf
k}}\sqrt{1+\frac{\tilde{H}^xH^x}{H^zH^z}}\coth (\frac{\beta\epsilon_{\bf
k}}{2})=0.
\label{5.39}
\end{equation}
The only difference from the corresponding equations of reference \cite{FJK00}
is that the
square root cannot be taken out of the sum (integral over the first Brillouin
zone) because of the momentum dependence of $H^x$ coming from the exchange
anisotropy. If the latter is zero as in reference \cite{FJK00},
only
the single-ion anisotropy survives and one recovers the original equations.

The explicit derivations above are done for pedagogical reasons, but they
would not have been necessary because
one obtains by subtracting equation (\ref{5.21}) from equation (\ref{5.22})
\begin{eqnarray}
& &2\tilde{H}^xH^xC_{\bf k}^{z,mn}-\tilde{H}^xH^zC_{\bf k}^{+,mn}
-\tilde{H}^xH^zC_{\bf k}^{-,mn}=\nonumber\\
& &\frac{1}{2}(A_{\eta=-1}^{-,mn}-A_{\eta=-1}^{+,mn})
\epsilon_{\bf k}\tilde{H}^x\coth(\frac{\beta\epsilon_{k}}{2})
+\frac{1}{2}(A_{\eta=-1}^{+,mn}+A_{\eta=-1}^{-,mn})\epsilon_{\bf
k}^2\frac{\tilde{H}^x}{H^z},
\end{eqnarray}
which corresponds to equation (27) of reference \cite{FJK00} which is the
starting point  for deriving the equations for the moments explicitly.

For the treatment of multilayers we have to use the eigenvector method as
mentioned above.
The essential features are as follows.
One starts with a transformation, which diagonalizes the $\Gamma$-matrix of
equation (\ref{5.8})
\begin{equation}
{\bf L\Gamma R}={\bf \Omega},
\label{5.40}
\end{equation}
where ${\bf \Omega}$ is a diagonal matrix with eigenvalues $\omega_{\tau}$
($\tau=1,..., 3N$), and the transformation matrix {\bf R} and its inverse
${\bf R}^{-1}={\bf L}$ are obtained from the right eigenvectors of
${\bf \Gamma}$
as columns and from the left eigenvectors as rows, respectively. These matrices
are normalized to unity: {\bf RL}={\bf LR}={\bf 1}.

Multiplying the equation of motion (\ref{5.8}) from the left by {\bf L} and
inserting {\bf 1}={\bf RL} one finds
\begin{equation}
(\omega{\bf 1}-{\bf \Omega}){\bf L}{\bf G}_\eta^{mn}={\bf LA}_\eta^{mn}.
\label{5.41}
\end{equation}
Defining ${\cal G}_\eta^{mn}={\bf LG}_\eta^{mn}$
and ${\cal A}_\eta^{mn}={\bf LA}_\eta^{mn}$ one obtains
\begin{equation}
(\omega {\bf 1}-{\bf \Omega}){\cal G}_\eta^{mn}={\cal A}_\eta^{mn}.
\label{5.42}
\end{equation}
${\cal G}_\eta^{mn}$ is a new vector of Green's functions, each component
$\tau$ of which has but a single pole
\begin{equation}
{\cal G}_\eta^{mn,\tau}=\frac{{\cal A}_\eta^{mn,\tau}}{\omega-\omega_\tau}\ .
\label{5.43}
\end{equation}
This is the important point and allows application of the spectral theorem to
each component separately. This gives with ${\cal C}^{mn}={\bf L}{\bf C}^{mn}$
\begin{equation}
{\cal C}^{mn,\tau}=\frac{{\cal
A}_{\eta}^{mn,\tau}}{e^{\beta\omega_\tau}+\eta}+\frac{1}{2}(1-\eta)\frac{1}{2}
\lim_{\omega\rightarrow 0}\frac{\omega{\cal
A}_{\eta=+1}^{mn,\tau}}{\omega-\omega_\tau} .
\label{5.44}
\end{equation}
In reference \cite{FJKE00} we used the commutator ($\eta=-1$).
Here, we proceed with the anticommutator ($\eta=+1$), so that the second
term in equation (\ref{5.44}) is zero and one obtains the original
correlation vector ${\bf C}^{mn}$ by multiplying ${\cal C}^{mn}$ from the left
with {\bf R}; i.e.
\begin{equation}
{\bf C}^{mn}={\bf R}{\bf {\cal E}}{\bf L}{\bf A}_{\eta=+1}^{mn},
\label{5.45}
\end{equation}
where ${\bf {\cal E}}$ is a diagonal matrix with matrix elements
${\cal E}_{ij}=\delta_{ij}(e^{\beta\omega_i}+1)^{-1}$.
With the relation (\ref{5.20}) we find
\begin{equation}
{\bf C}^{mn}={\bf R}{\bf {\cal E}}{\bf L}({\bf A}_{\eta=-1}^{mn}+2{\bf
C}^{mn})\ , \label{5.46}
\end{equation}
or
\begin{equation}
{\bf C}^{mn}=(1-2{\bf R{\cal E} L})^{-1}{\bf R{\cal E} LA}_{\eta=-1}^{mn}.
\label{5.47}
\end{equation}
For the monolayer it can be shown explicitly that the eigenvector method yields
the equations (\ref{5.21},\ref{5.22},\ref{5.23}) derived before.
In this case the eigenvectors by which the transformation matrices
{\bf R } and {\bf L} are constructed can be given explicitly. They are
\begin{equation}
{\bf R}=\left(\begin{array}{ccc}
\frac{H^x}{H^z} &\frac{-(\epsilon_{\bf
k}+H^z)}{\tilde{H}^x}&\frac{(\epsilon_{\bf k}-H^z)}{\tilde{H}^x}\\
\frac{H^x}{H^z} &\frac{(\epsilon_{\bf
k}-H^z)}{\tilde{H}^x}&\frac{-(\epsilon_{\bf k}+H^z)}{\tilde{H}^x}\\
1&1&1
\end{array}\right)\ ,
\label{5.48}
\end{equation}
and
\begin{equation}
{\bf L}=\frac{1}{4\epsilon_{\bf k}^2}\left(
\begin{array}{ccc}
2\tilde{H^x}H^z&2\tilde{H^x}H^z&4H^zH^z\\
-(\epsilon_{\bf k}+H^z)\tilde{H}^x&(\epsilon_{\bf
k}-H^z)\tilde{H}^x&2H^x\tilde{H}^x\\
(\epsilon_{\bf k}-H^z)\tilde{H}^x&-(\epsilon_{\bf
k}+H^z)\tilde{H}^x&2H^x\tilde{H}^x
\end{array}\right)\ .
\label{5.49}
\end{equation}
Putting this into equation (\ref{5.46}) yields
equations (\ref{5.21},\ref{5.22},\ref{5.23}).

In order to obtain the correlations in real space, equation (\ref{5.47}) has to
be Fourier transformed and the resulting integral equation has to be solved
self-consistently.
By inspecting the expressions for the
monolayer, one can show that the inverse
$(1-2{\bf R}{\bf \cal{E}}{\bf L})^{-1}$ does
not exist . Therefore this equation cannot furnish the solution.
However, one can show that the formulation with the anticommutator
relation can be transformed into the result for the commutator relation:
${\bf \cal E}$ is a diagonal matrix with
${\cal E}_{ij}=\delta_{ij}(e^{\beta\omega_i}+1)^{-1}$. With the relation
$({\bf R{\cal E}L})^{-1}={\bf L}^{-1}({\bf {\cal E}})^{-1}{\bf R}^{-1}={\bf
R}{\bf {\cal E}}^{-1}{\bf L}$ one obtains from equation (\ref{5.47})
\begin{eqnarray}
{\bf C}&=&({\bf R}({\bf 1}-2{\bf \cal E}){\bf L})^{-1}{\bf R{\cal
E}LA}\nonumber\\
 &=&{\bf R}(\bf{1}-2{\bf {\cal E}})^{-1}{\bf {\cal E}}{\bf LA}\nonumber\\
&=& {\bf R\tilde{\cal E}LA},
\label{5.50}
\end{eqnarray}
where $\tilde{\cal E}_{ij}=\delta_{ij}\frac{{\cal E}_{ii}}{1-2{\cal
E}_{ii}}=\delta_{ij}(e^{\beta\omega_i}-1)^{-1}$,
which still is of no use because it diverges for $\omega_i=0$.

In reference \cite{FJKE00}, it was shown that the matrix ${\bf R}_0{\bf
L}_0$, where the index refers to the eigenvectors with eigenvalue zero, is a
projection operator onto the zero eigenvalue space.
The situation can then be remedied by projecting  equation (\ref{5.50}) onto
the non-zero eigenvalue space
with ${\bf1}-{\bf R}_0{\bf L}_0$, which leads to the commutator expression for
the correlations
of ref. \cite{FJKE00}
\begin{equation}
{\bf C}={\bf R\tilde{\cal E}_0LA}+ {\bf R}_0{\bf L}_0{\bf C},
\label{5.47a}
\end{equation}
where $\tilde{\cal E}_0$ is equal to $\tilde{\cal E}$ in which the diagonal
elements corresponding to $\omega=0$ have been set to zero.

The problem one is confronted with in applying the eigenvector method to this
equation is that the
the exchange anisotropy introduces
a momentum dependence in
the projection operator. Then the projector cannot be taken out of the integral
when
a Fourier transform to real space is performed
as in the case of the Anderson Callen decoupling of the
single-ion anisotropy only.
The way out is to eliminate the projector by a transformation in one component
of equation (\ref{5.47a}), which is sufficient to  establish the integral
equations of the eigenvector method. This procedure was inspired by the
elimination of the disturbing term in equations (\ref{5.21},\ref{5.22}).

The adequate transformation  is found to be
\begin{equation}
{\bf T}^{-1}=\frac{1}{2}\left(\begin{array}{ccc}
1 & 1 & 0 \\
-1& 1 & 0  \\
0 & 0 & 2
\end{array}\right)\ \ \ \ \ \ \
{\bf T}=\left(\begin{array}{ccc}
1 & -1 & 0\\
1 & 1 & 0 \\
0 & 0& 1
\end{array}\right)
\label{5.53a}
\end{equation}
with ${\bf T}^{-1}{\bf T}={\bf 1}$.

Applying this transformation to equation (\ref{5.47a})
\begin{equation}
{\bf T}^{-1}{\bf C}={\bf T}^{-1}{\bf R\tilde{\cal E}_0LTT}^{-1}{\bf A}+{\bf
T}^{-1}{\bf R}_0{\bf L}_0{\bf TT}^{-1}{\bf C}
\label{5.55}
\end{equation}
and inserting the eigenvectors (\ref{5.48},\ref{5.49}) transforms the
second component of the vector
${\bf T}^{-1}{\bf R}_0{\bf L}_0{\bf TT}^{-1}{\bf C}$ to zero, and one recovers
equation (\ref{5.25}) from the second row of the
transformed equation (\ref{5.55}), from which, together with the regularity
conditions, the integral equations for the correlations for each ($m,n$)-pair are
obtained.

The eigenvector method is then
immediately generalized to the case of $N$ layers by
applying the transformation to equation (\ref{5.47a}) read as a
$3N$-dimensional problem, thus constructing
$3N\times 3N$-matrices with sub-matrices formed with the
transformation (\ref{5.53a}) on the diagonals.
\newpage
\subsection*{3. Numerical results}

In this section we discuss results of calculations for a square lattice
including the exchange anisotropy.
The integral equations (\ref{5.55}) for determining the correlations
together with the regularity conditions as derived in Appendix A of this
paper are solved self-consistently by the curve
following method, which we described in detail in Appendix A of reference
\cite{FKS02}.
A comparison with the results for the single-ion anisotropy as used in
our previous work \cite{FJK00,FJKE00} is effected by
fixing the strength of the exchange anisotropy such
that the Curie temperature of a square monolayer with spin $S=1$ agrees
approximately with
that of a single-ion anisotropy with strength $K_2=1$, which was used in
most of the calculations of our previous work. The exchange interaction
strength, $J=100$, and the dipole coupling strength, $g=0.018$ (corresponding
to the case of Ni), are taken to be the same in both kinds of calculations.
The exchange anisotropy coupling strength turns out to be D=0.7. In order to
compare results
for different spin values all parameters are scaled as $J\rightarrow J/S(S+1)$,
$g\rightarrow g/S(S+1)$, $D\rightarrow D/S(S+1)$, $K_2\rightarrow K_2/S(S-1/2)$
and if a magnetic field is applied, ${\bf B}\rightarrow {\bf B}/S$.

\begin{figure}[htb]
\begin{center}
\protect
\includegraphics*[bb = 80  90 540 700,
angle=-90,clip=true,width=10cm]{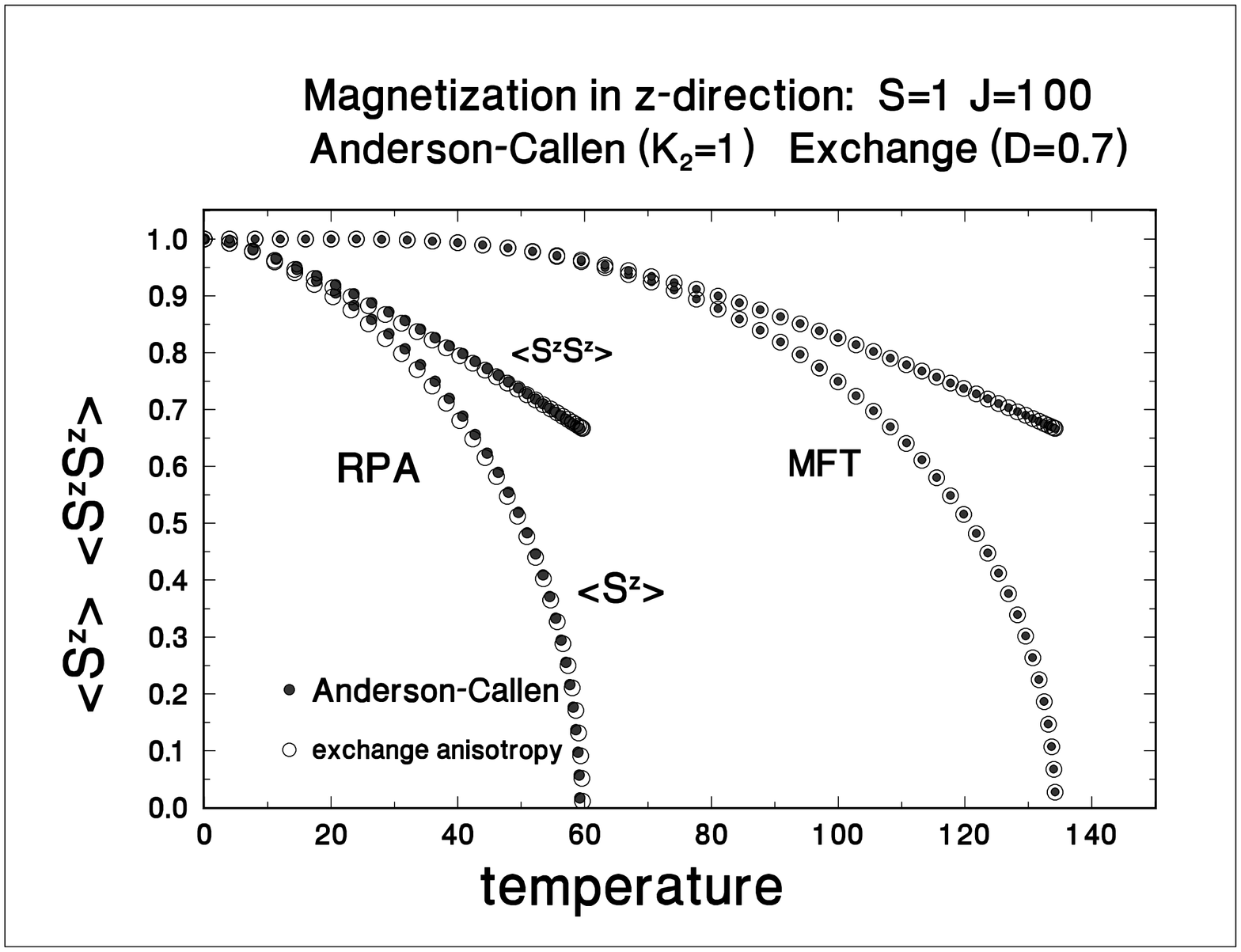}
\protect
\caption{The magnetization $\la S^z\ra$ and $\la S^zS^z\ra$ of a ferromagnetic
spin $S=1$ Heisenberg monolayer for a square lattice are shown as
functions of the temperature (no magnetic field). Comparison
is made between Green's function (RPA) calculations using the exchange
anisotropy ($D=0.7$, open circles) and the single-ion anisotropy ($K_2=1$,
solid dots) with Anderson-Callen decoupling. The corresponding results of
mean field (MFT)  calculations are also shown. }
\label{fig1}
\end{center}
\end{figure}

In figure 1 we display the magnetization $\la S^z\ra$ and its second moment
$\la S^zS^z\ra$ as functions of the temperature for a $S=1$ monolayer using
Green's function theory (denoted as RPA). Using a value of $D=0.7$ as
the strength of the exchange anisotropy yields nearly the same curve as for an
Anderson-Callen decoupling of the single-ion anisotropy with a strength of
$K_2=1$. The results for the corresponding mean field (MFT)
calculations also practically coincide with each other. As
is well known and also discussed in our previous work, the neglect of magnons
results in
a Curie temperature which is more than a factor of two larger than that
obtained by including
magnon excitations. This difference for a monolayer is much larger than the
corresponding difference for bulk ferromagnets.
\begin{figure}[htb]
\begin{center}
\protect
\includegraphics*[bb = 80  90 540 717,
angle=-90,clip=true,width=10cm]{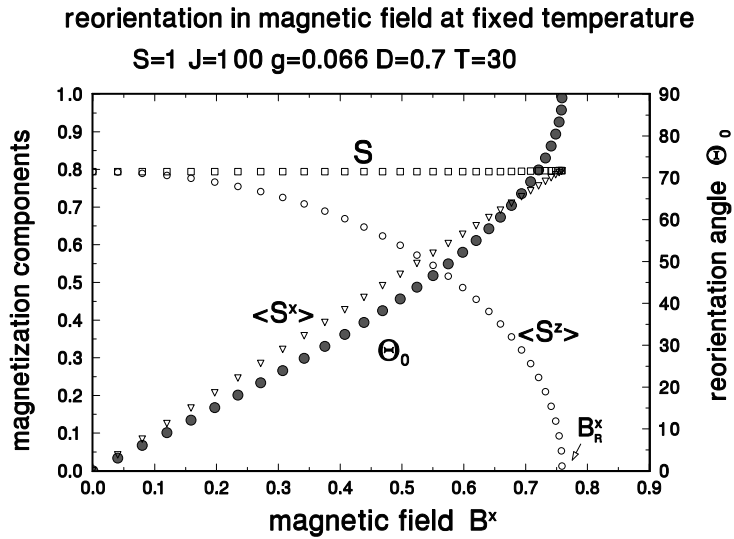}
\protect
\caption{The components of the magnetization $\la S^z\ra$ and
$\la S^x\ra$ and the absolute value $S$ at a fixed temperature $T=30$
as function of
an external magnetic field in the $x$-direction, $B^x$, are shown for a
ferromagnetic spin $S=1$ Heisenberg monolayer for a square lattice.
Also shown are the equilibrium reorientation angle $\theta_0$ and the critical
reorientation field, $B^x_R$, at which in-plane orientation is reached.}
\label{fig2}
\end{center}
\end{figure}

In figure 2 we show  the reorientation of the magnetization at a fixed
temperature due to a transverse field in $x$-direction, $B^x$. The magnetic
field in $z$-direction is set to zero, $B^z=0$.
In this case, the strength of the dipole coupling is chosen to be $g=0.066$ (a
value corresponding to Co).
As a function of the external field, the $x$-component of the magnetization
$<S^x>$ rises linearly, whereas its $z$-component $<S^z>$ falls to zero,
where
the in-plane magnetization is reached ($\theta_0=90^o$). The absolute value
$S=\sqrt{<S^x>^2+<S^z>^2}$ remains constant, as it should do, and is also
shown in the figure.
\begin{figure}[htb]
\begin{center}
\protect
\includegraphics*[bb = 80  90 540 700,
angle=-90,clip=true,width=10cm]{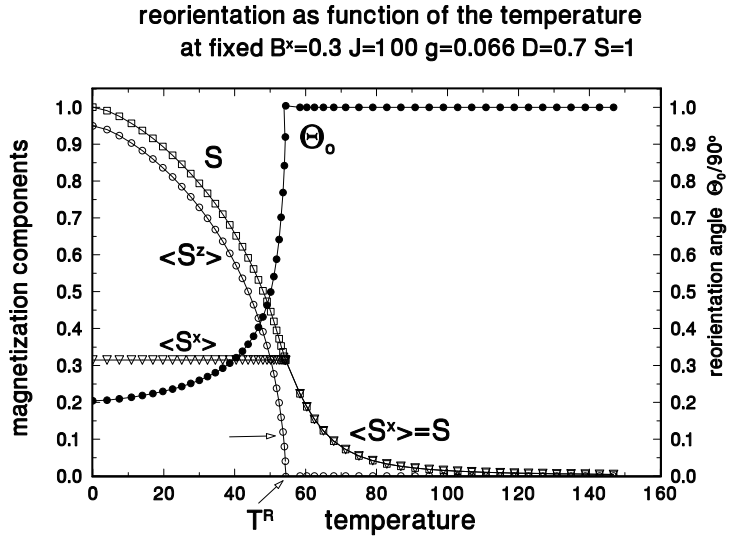}
\protect
\caption{
The components of the magnetization $\la S^z\ra$ and
$\la S^x\ra$ and its absolute value $S$ for a fixed magnetic field
$B^x=0.3$ as function of
the temperature are shown for a ferromagnetic
spin $S=1$ Heisenberg monolayer for a square lattice.
Also shown are the equilibrium reorientation angle $\theta_0$ and the critical
reorientation temperature, $T_R$, at which in-plane orientation is reached.
The small horizontal arrow indicates the value of $\la S^z\ra$ below which
complex eigenvalues occur.} \label{fig3} \end{center}
\end{figure}

In figure 3, we show the magnetization components as functions of the
temperature for the same parameters as in figure 2
at a constant external field, $B^x=0.3$. The component $\la S^x\ra$ stays
constant until the component $\la S^z\ra$ has dropped to zero, and an in-plane
magnetization ($\theta_0=90^o$) is reached. The fact that $\la S^x\ra$ is
constant for
temperatures below the reorientation temperature, $T<T_R$, can be understood
from the regularity condition (\ref{5.35}), which, for a monolayer with
scaled parameters ($\la S^z\ra$ drops out of the expression if $B^z=0$),
is $\la S^x\ra/S=\frac{B^x/S}{(4D-3gT^0)/S(S+1)}$. This expression
restricts the range  of parameters for the validity of our approach because
$\la S^x\ra/S$ must be $\leq 1$. Above the reorientation temperature,
$\la S^z\ra$ remains zero whereas $\la S^x\ra$  remains finite
because of
the field in the $x$-direction and decreases slowly with increasing
temperature. The absolute value
of the magnetization $S$ is also shown; above $T_R$ one has $S=\la S^x\ra$.

With the chosen parameters one observes a novel feature in the calculations
which is connected with the introduction of the exchange anisotropy
{\em together}
with the dipole coupling. In this case, the eigenvalues of equation
(\ref{5.17})
become complex above a certain temperature, i.e. below a certain value of
$\la S^z\ra $. This behaviour occurs quite naturally in the theory.
Because the $\Gamma$-matrix of equation (\ref{5.8}) is real, its eigenvalues
and
eigenvectors, if complex, occur pairwise as complex conjugates, and the term
${\bf R}\tilde{\cal E}{\bf L}$ in the equation (\ref{5.55}) is real, so that
one has to do with a real integral equation.
The
complex eigenvalues and vectors have to be taken seriously, and are necessary
for obtaining the results
of figure 3. The complex eigenvalues are connected with the additional term in
$H_i^\pm$ of equation (\ref{5.12}), which comes from the exchange anisotropy.
This can be seen analytically by considering the dispersion relation
(\ref{5.17}), which shows that the optimal condition for
the occurrence
of complex eigenvalues is at ${\bf k}=0$ ( thus $\gamma_{\bf k}=4$). With q=4,
$B^z=0$, $K_2=0$ one has for the monolayer from equation (\ref{5.17})
\begin{equation}
\epsilon_{{\bf k}=0}^2=(4D-2gT^0)^2\la S^z\ra^2-(gT^0\la S^x\ra+B^x)(\la
S^x\ra(4D-gT^0)-B^x)
\end{equation}
Complex eigenvalues occur if the second term is larger than the first term.
Using
the regularity condition (\ref{5.36}) for
$\la S^x\ra$, one obtains (using
scaled parameters for $S=1$) a condition for the occurrence of complex
eigenvalues,
\begin{equation}
\la S^z\ra\  <\  \frac{2B^x}{(4D-3gT^0)}\sqrt{\frac{2gT^0}{(4D-2gT^0)}}.
\end{equation}
This yields complex eigenvalues for $\la S^z\ra<0.1157$ at $T\simeq 53$ for
the parameters used in figure 3. This is confirmed in the numerical
calculations, where of course complex eigenvectors also occur at finite
${\bf k}$. Enlarging $B^x$ and/or g increases the range of complex eigenvalues.
 No complex eigenvalues occur if the dipole coupling is set equal to
zero.
In our previous work \cite{FJKE00}, where we used the single-ion
anisotropy together with the dipole coupling, complex eigenvalues
never occured, which can be understood from the structure of the dispersion
relation (\ref{5.17}) by putting the exchange anisotropy terms equal to zero.

\begin{figure}[htb]
\begin{center}
\protect
\includegraphics*[bb = 80  90 540 700,
angle=-90,clip=true,width=10.5cm]{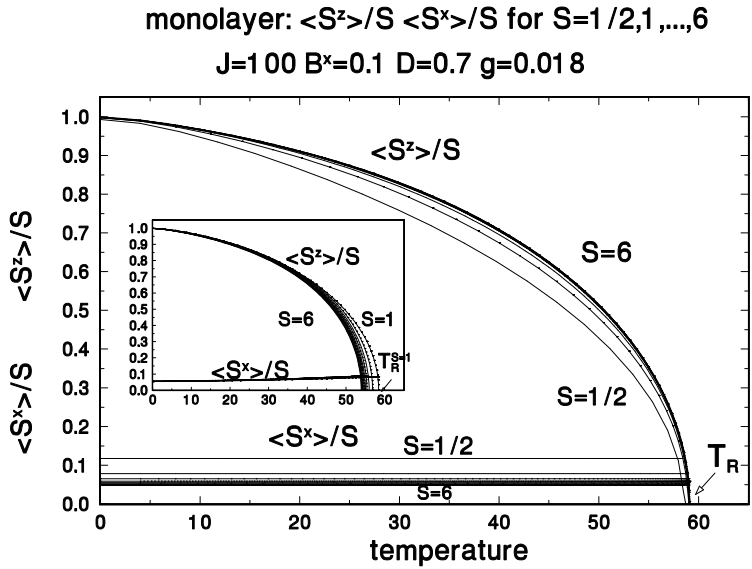}
\protect
\caption{Normalized magnetization curves for $\la S^z\ra/S$ and
$\la S^x\ra/S$ for a monolayer with spin values between $S=1/2$ and $S=6$
shown as functions of the temperature using the exchange anisotropy.
The corresponding results obtained with a
single-ion anisotropy ($K_2=1$) are shown in the inset (from reference
\cite{FJKE00}). } \label{fig4}
\end{center}
\end{figure}
In Fig.4 we display the normalized magnetizations $\la S^z\ra/S$ of a
monolayer as functions of the temperature $T$ for all half-integral and
integral spin values ranging from $S=1/2$ to $S=6$. In the case of the exchange
anisotropy, the reorientation temperature $T_R$ is practically the same for all
spin values. One observes a spin dependence of the magnetization curves which
decreases with increasing spin. The curves, however, saturate very quickly;
on the scale of the figure, one cannot distinguish the curves between $S=2$ and
$S=6$.
When using the single-ion anisotropy (the results are shown as an
inset) there is a difference in the reorientation temperatures, but the
magnetization curves again saturate very quickly (approaching the
classical limit) but in the opposite direction. In both cases, the values for
$\la S^x\ra/S$ remain
rather small owing to the application of the small field in
the $x$-direction $B^x=0.1$.

\begin{figure}[htb]
\begin{center}
\protect
\includegraphics*[bb = 80  90 540 700,
angle=-90,clip=true,width=14cm]{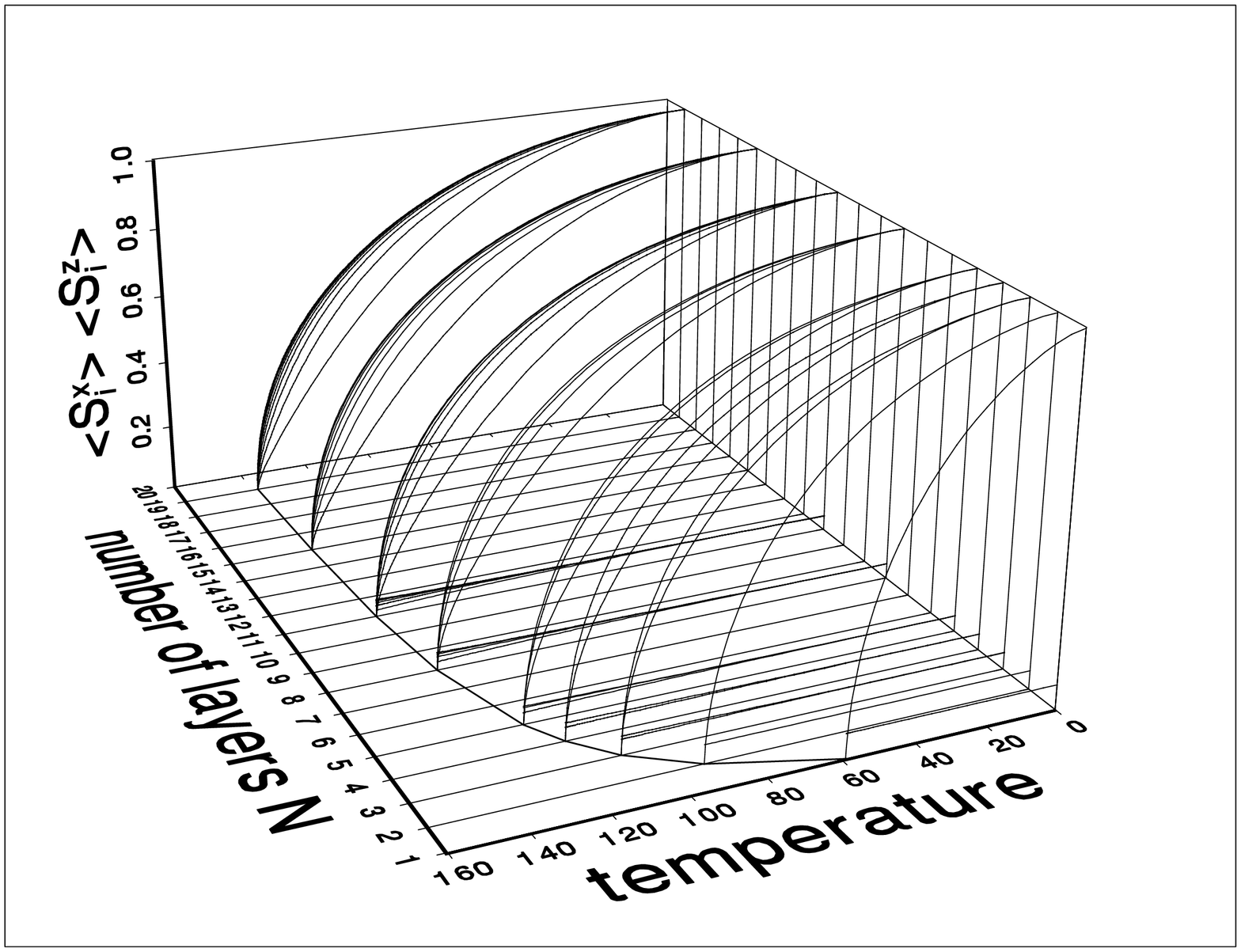}
\protect
\caption{
Sublayer magnetization components $\la S^z_i\ra$ and $\la S^x_i\ra$ as
functions of the temperature for spin $S=1$ films with $N$ layers calculated
with the exchange anisotropy. The
reorientation  temperatures $T_R^N$ can be read from the curve in the N-T
plane.}
\label{fig5} \end{center}
\end{figure}
In figure 5, we display the
sublayer magnetization components $\la S^z_i\ra$ and $\la S^x_i\ra$ as
functions of the temperature for spin $S=1$ films with thicknesses ranging
from $N=1$ to $N=19$ layers calculated with the exchange anisotropy using the
same parameters as in figure 4.
The reorientation  temperatures $T_R^N$ can be read  from the curve in the N-T
plane, which approaches the bulk value with increasing layer thickness.
Similar results were obtained in reference \cite{FJKE00} when using
the single-ion anisotropy. To have a direct comparison of the reorientation
temperatures calculated with the exchange
anisotropy ($D=0.7$) and the single-ion anisotropy ($K_2=1$), we display figure
6 . Because the
exchange anisotropy parameter was fitted to give the same result for the Curie
temperature as a calculation with the anisotropy parameter for the monolayer,
the corresponding reorientation temperatures practically coincide in this
case, whereas
the reorientation temperatures turn out to be slightly higher for the exchange
anisotropy
calculations with increasing film thickness. The saturation towards the
bulk limit follows the same trend in both cases.

\begin{figure}[htb]
\begin{center}
\protect
\includegraphics*[bb = 80  90 540 700,
angle=-90,clip=true,width=10cm]{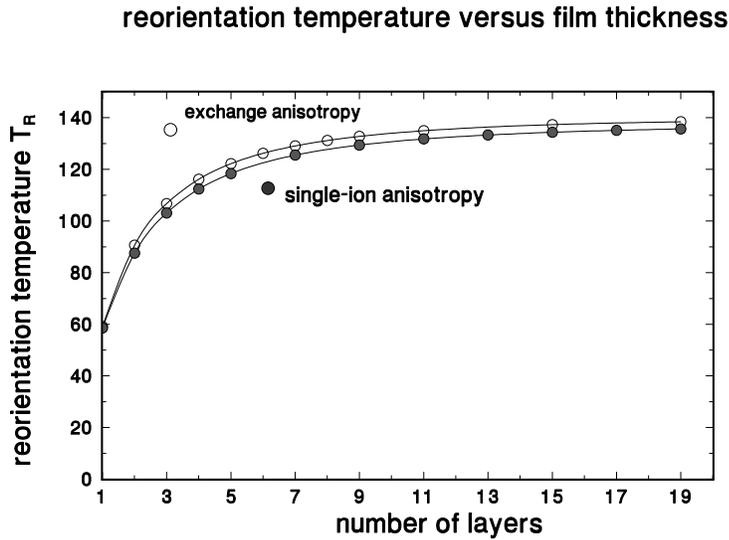}
\protect
\caption{Reorientation temperature as a function of the film thickness
displayed for results with the exchange anisotropy and the Anderson-Callen
treatment of the single-ion anisotropy (see reference \cite{FJKE00} for the
latter).}
\label{fig6}
\end{center}
\end{figure}
\begin{figure}[htb]
\begin{center}
\protect
\includegraphics*[bb = 80  90 540 700,
angle=-90,clip=true,width=10cm]{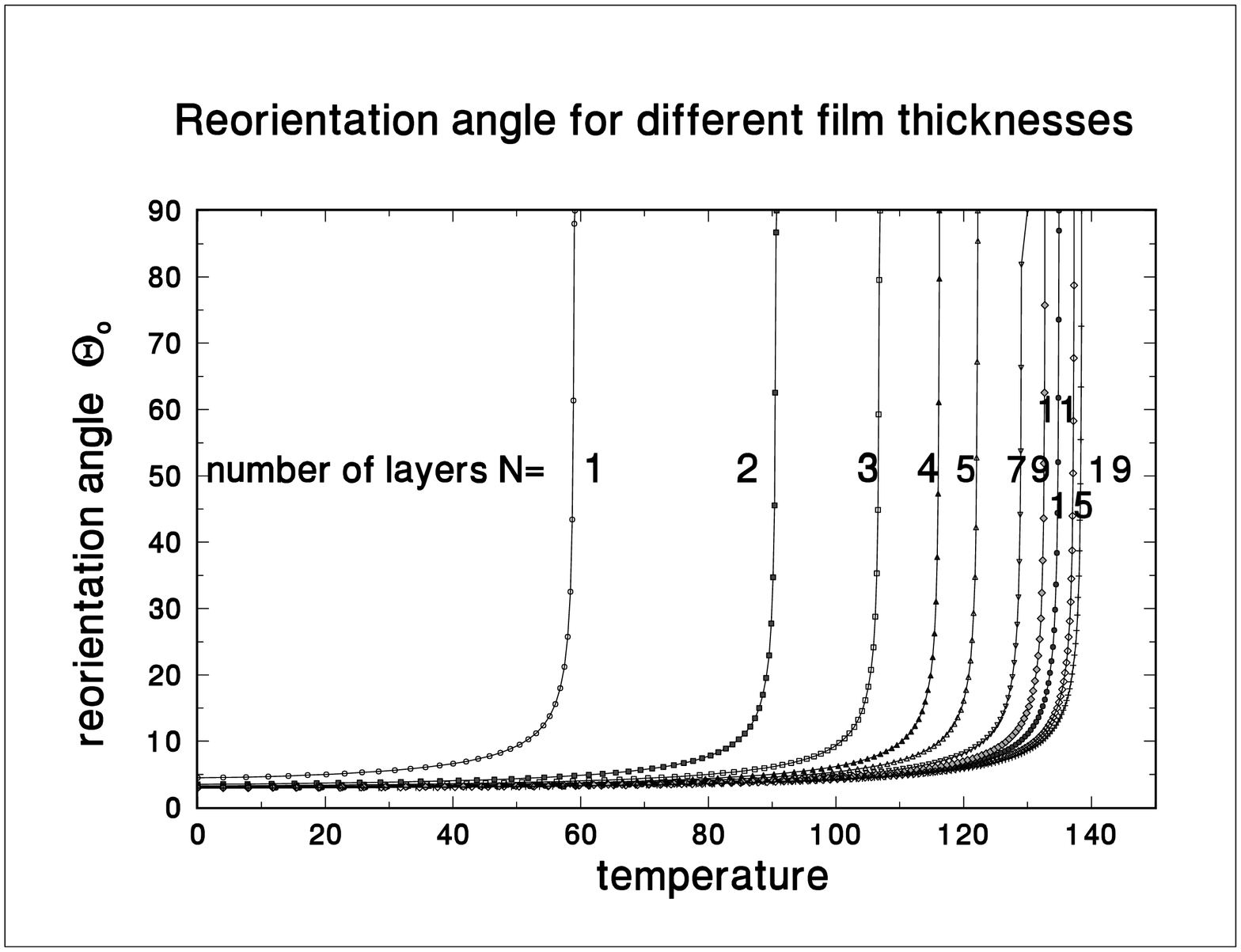}
\protect
\caption{ Average equilibrium reorientation angle for different film
thicknesses (number of layers N=1,..., 5, 7, 9, 11, 15, 19) shown as a
function of the temperature. This is a result of calculations using the
exchange anisotropy and the parameters are the same as in the previous figures.
} \label{fig7} \end{center}
\end{figure}
\newpage

In figure 7 we show the average reorientation angle $\theta_0(N,T)$ as a
function of the temperature for films with increasing film thickness, where we
define
\begin{equation}
\theta_0(N,T)=\arctan\frac{
\frac{1}{N}\sum_{i=1}^N \la S^z_i\ra}
{\frac{1}{N}\sum_{i=1}^N \la S^z_i\ra}\ .
\end{equation}
The curves show the same saturation behaviour in the bulk limit as
already seen in the previous figures.

\newpage
\newpage
\subsection*{4.Conclusions}
In the present paper, we have included the exchange anisotropy in our many-body
Green's function description of thin ferromagnetic films. A few non-trivial
changes in the general formalism had to be implemented because of additional
momentum dependencies when the exchange anisotropy is included.
The momentum dependence of the projector onto the zero eigenvalue space has
been eliminated by an appropriate transformation. A novel feature is the
appearance of complex eigenvalues and eigenvectors of the non-symmetric matrix
governing the equation of motion for the Green's functions when including the
exchange anisotropy and the dipole coupling together. The complex eigenvalues
occur quite naturally in the theory and have to be taken seriously. They are
necessary to obtain correct results above the temperature where the
magnetization $\la S^z\ra$ drops below a certain value.

The physical quantities calculated with the exchange anisotropy or with the
single-ion anisotropy are fairly similar when the exchange anisotropy strength
is fitted in such a way that it gives the same Curie temperature for a spin
$S=1$ Heisenberg monolayer as that calculated with the single-ion anisotropy
parameter of a certain strength, the rest of the parameters such as the
exchange interaction strength and the dipole coupling strength being the same.

In the present paper, all intra- and interlayer coupling parameters have been
taken to be the same but the computer program is written in such a way that
they can easily be chosen differently. We have also shown only examples for a
monolayer
with spins in the range between $S=1$ and $S=6$ and films with $N$ layers for
spin $S=1$. It is only a question of computer time to make calculations for
films with other ($S,N$) combinations.

\subsection*{Acknowledgement}
We are indebted to P.J. Jensen for very useful discussions.

\newpage

\subsubsection*{Appendix A: Treating ${\bf S\geq 1}$}
In this Appendix we show how the regularity conditions
can be deduced for general spin
quantum numbers $S$ and for multilayers.
From the definitions (\ref{5.12}) we see from (\ref{5.35})
that
\begin{equation}
\frac{\tilde{H}^{x}_i}{H^z_i}=\frac{B^x_i}{Z_i}.
\label{A1}
\end{equation}
The regularity conditions (\ref{5.34}) can therefore be written for general
$m,n$ in the form for each layer $i$
\begin{equation}
-2Z_iA_{-1,i}^{z,mn}=A_{-1,i}^{+,mn}B^x_i+A_{-1,i}^{-,mn}B^x_i\ .
\label{A2}
\end{equation}
For the calculation of the correlations for higher spin we use equation
(\ref{5.25}) generalized to the multilayer case. We leave out the layer index
$i$ in all formulas which follow.
\begin{eqnarray}
& &\la (S^z)^m(S^-)^nS^+\ra-\la (S^z)^m(S^-)^nS^-\ra
-\frac{1}{2}(A_{\eta=-1}^{-,mn}-A_{\eta=-1}^{+,mn})=\nonumber\\
& &\frac{1}{2}(A_{\eta
=-1}^{-,mn}+A_{\eta=-1}^{+,mn})
\frac{1}{N}\sum_{\bf k}\frac{\epsilon_{\bf
k}}{H^z}\coth(\frac{\beta\epsilon_{\bf k}}{2}) .
\label{A3}
\end{eqnarray}
We express all correlation functions occuring in this equation
in a standard form
where all powers of $S^z$ are written to the left of the powers of $S^-$:
\begin{equation}
C(m,n)=\langle(S^z)^m(S^-)^n\rangle.
\label{A.4}
\end{equation}
Then, with the relations $[S^z,(S^-)^n]_-=-n(S^-)^n$ and
$S^-S^+=S(S+1)-S^z-(S^z)^2$, we find that
\begin{eqnarray}
\langle (S^z)^m(S^-)^nS^z\rangle &=&nC(m,n)+C(m+1,n)\ ,\nonumber\\
\langle
(S^z)^m(S^-)^nS^+\rangle&=&\Big(S(S+1)-n(n-1)\Big)C(m,n-1)-(2n-1)C(m+1,n-1)
\nonumber\\ & &-C(m+2,n-1)\ ,\nonumber\\
 \langle (S^z)^m(S^-)^nS^-\rangle&=&C(m,n+1)\ .
\label{A5}
\end{eqnarray}
The commutators can also be expressed in terms of the
$C(m,n)$ using the binomial series
\begin{eqnarray}
A_{-1}^{z,mn}&=&-nC(m,n)\ ,\nonumber\\
A_{-1}^{+,mn}&=&\langle\Big[\Big((S^z-1)^m-(S^z)^m\Big)S^-S^++2S^z(S^z-1)^m
+(n-1)(n+2S^z)(S^z)^m
\Big](S^-)^{n-1}\rangle \nonumber \\
& & =S(S+1)\sum_{i=1}^m \left( \begin{array}{c} m\\ i \end{array} \right)
(-1)^iC(m-i,n-1)+(2n+m)C(m+1,n-1)\nonumber\\
& &+\sum_{i=2}^{m+1} \left( \begin{array}{c} m+1\\ i \end{array} \right)
(-1)^{i+1}C(m+2-i,n-1)+n(n-1)C(m,n-1)\ ,
\nonumber \\
A_{-1}^{-,mn}&=&\langle \big[ (S^z+1)^m-(S^z)^m \big](S^-)^{n+1}\rangle
=\sum_{i=1}^m \left( \begin{array}{c} m\\ i \end{array} \right)C(m-i,n+1)\ .
\label{A6}
\end{eqnarray}
Now by putting equation (\ref{A6}) into equation (\ref{A2}) the regularity
conditions for all $m$ and $n$ can be written in terms of correlations defined
in the standard form:
\begin{eqnarray}
& &2ZnC(m,n)=B^x\Big[
S(S+1)\sum_{i=1}^m \left( \begin{array}{c} m\\ i \end{array} \right)
(-1)^iC(m-i,n-1)\nonumber \\ & &+(2n+m)C(m+1,n-1)
+\sum_{i=2}^{m+1} \left( \begin{array}{c} m+1\\ i \end{array} \right)
(-1)^{i+1}C(m+2-i,n-1)\nonumber \\
& & +n(n-1)C(m,n-1) \Big]
+B^x\sum_{i=1}^m \left( \begin{array}{c} m\\ i \end{array}
\right)C(m-i,n+1)\ \ .
\label{A.7}
\end{eqnarray}
For a given spin $S$, this set of linear equations for the correlations has to
be solved
for all $m+n\leq 2S+1$. The solutions have to be put via equations
(\ref{A5}) together with (\ref{A6})
into equations (\ref{A3}), thus leading to a set of $2S$ equations for the
moments $\la(S^z)^p\ra$ (p=1,\ldots,2S), which have to be
solved self-consistently.
The highest moment $\la (S^z)^{2S+1}\ra$ has
been eliminated in favour of the lower ones through the relation
$\prod_{M_S}(S^z-M_S)=0$.

%For the multilayer the correlations and the quantity $Z$ carry a layer index
%
%\begin{eqnarray}
%Z_i&=&B^z_i+D_{ii}q\la S^z_i\ra
%+(J_{i,i+1}+D_{i,i+1})\la S_{i+1}^{z}\ra+(J_{i,i-1}+D_{i,i-1})\la
%S_{i-1}^{z}\ra\nonumber\\
%& &+K_{2,i}2\la S_i^z\ra
%\Big(1-\frac{1}{2S^2}[S(S+1)-\la S_i^zS_i^z\ra]\Big)\ .
%\end{eqnarray}
%


\newpage\begin{thebibliography}{99}

%1
\bibitem{Mos94} A. Moschel, K.D. Usadel, Phys. Rev. B {\bf 49}, 12868 (1994);
P.J. Jensen, K.H. Bennemann, in {\em 'Magnetism and Electronic Correlations in
Local-Moment Systems: Rare-Earth Elements and Compounds'}, ed. M. Donath, P.A.
Dowben, and W. Nolting, (World Scientific, 1998), p. 113-141.
%2
\bibitem{Jen96} P.J. Jensen, K.H. Bennemann, Solid State Comm. {\bf 100}, 585
(1996), ibid. {\bf 105},577 (1998); A. Hucht, K.D. Usadel, Phys. Rev. B {\bf
55}, 12309 (1997).
%3
\bibitem{EFJK99} A. Ecker, P. Fr\"obrich, P.J. Jensen, P.J. Kuntz,
J. Phys.: Condens. Matter {\bf 11}, 1557 (1999).
%4
\bibitem{Tya59} S.V. Tyablikov, Ukr. Mat. Zh. {\bf 11}, 289 (1959).
%5
\bibitem{FJK00} P. Fr\"obrich, P.J. Jensen, P.J. Kuntz, Eur. Phys. J. B
{\bf13}, 477 (2000).
%6
\bibitem{AC64} F.B. Anderson, H.B. Callen,  Phys. Rev. {\bf
136}, A1068 (1964).
%7
\bibitem{FKS02} P. Fr\"obrich, P.J. Kuntz, M. Saber, Ann. Phys. (Leipzig)
{\bf 11}, 387 (2002).
%8
\bibitem{HFKTJ02} P. Henelius, P. Fr\"obrich, P.J. Kuntz, C. Timm, P.J. Jensen,
Phys. Rev. B {\bf 66}, 094407 (2002).
%9
\bibitem{FJKE00} P. Fr\"obrich, P.J. Jensen, P.J. Kuntz, A. Ecker, Eur. Phys.
J. B {\bf18}, 579 (2000).
%10
\bibitem{Eric91} R.P. Erickson, D.L. Mills, Phys. Rev. B {\bf 44}, 11825
(1991).
%11
\bibitem{Guo00} Wenli Guo, L.P. Shi, D.L. Lin, Phys. Rev. B {\bf 62}, 14259
(2000).
%12
\bibitem{HLT99} L Hu, H. Li, R. Tao, Phys. Rev. B {\bf 60}, 10222 (1999).
%13
\bibitem{DGS02} M. Dantziger, B. Glinsmann, S. Scheffler, B. Zimmermann, P.J.
Jensen, Phys. Rev. B {\bf 66}, 094416 (2002).
%14
\bibitem{Lin67} M.E. Lines, Phys. Rev. {\bf156}, 534 (1967).
%15
\bibitem{Cal63} H.B. Callen, Phys. Rev. {\bf 130}, 890 (1963).
%
\bibitem{GHE01} W. Gasser, E. Heiner, and K. Elk, in 'Greensche Funktionen in
der Festk\"orper- und Vielteilchenphysik', Wiley-VHC, Berlin, 2001, Chapter
3.3.
% %
\end{thebibliography}
\end{document}